\RequirePackage{fix-cm}
\documentclass[smallextended]{svjour3}       
\smartqed  
\usepackage{graphicx}
\begin{document}

\title{T-bulge single photon quantum router with three level system}
\author{Lin Liu}
\institute{Lin Liu \at
              Key Laboratory of Low-Dimensional Quantum Structures
and Quantum Control of Ministry of Education, and Department of
Physics, Hunan Normal University, Changsha 410081, China \\
              North China Institute of Aerospace
Engineering, Langfang 065000, China \\
             \email{liulin@hunnu.edu.cn}       
        }
\date{Received: date / Accepted: date}
\maketitle

\begin{abstract}
This paper considers the transmission characteristics of incidenting from an infinte coupled-resonator waveguide (CRW) or a semi-infinite CRW, respectively. The Nth cavity of a semi-infinite CRW intersecting with an infinite CRW and a cascade three level system(CTLS) form a T-bulge structure. Due to symmetry breaking boundary, the maximum of transfer rate would reach unity when light incidenting from the semi-infinite CRW, while reach 0.5 for infinite CRW case. The position of the intersection have effect on the location of the extreme point of the transmission coefficients changing with couple strength. 4D figures are used to illustrate the transmission characteristics, which is more visable.

\end{abstract}

\section{Introduction}
\label{intro}
    Quantum networks composed of many nodes and channels, have many application in quantum communication. Quantum router, which acting as key component of quantum networks, can be achieved by the optical interactions of single photons and atoms, allowing the distribution of entanglement across the network~\cite{NAT453}. Measurement-device-independent quantum key can distribute over a 404 km optical fiber~\cite{PRL190501}. The coupling strength can be increased at frequencies near resonances in free space. The employment of a cavity can enhance the coupling strength further~\cite{PRL043807,PRA044304}.

    Zhou \emph{et al} have made good jobs on quantum router, utilising a controllable two-level system as a quantum switch for the coherent transport of a single photon~\cite{PRL100501}, a cyclic three-level system embedded in the junction of two infinite CRWs forming an X-type quantum router~\cite{PRL111(13)}, changing the cyclic type into a new type with an inversion center~\cite{PRA89(14)103805}. Then make alter both in the atom and the CRWs' structure, a two-level system (TLS) in a T-shaped waveguide made of an infinite CRW and a semi-infinite CRW~\cite{OE23(15)}.

    Now, changing into a CTLS couple with two CRWs, the Nth cavity of a semi-infinite CRW intersecting with an infinite CRW, let we see what happens to the transmission characteristics of incidenting from each CRW.

    This paper is organized as follows: In Sec. 2, the model used in this paper is introduced. In Sec. 3 and Sec. 4, the single-photon scattering process is studied for waves incident from different CRWs. Finally, we conclude with a brief summary of the results.
\section{The model}
\label{sec:2}

As shown in Fig. 1, coupled resonators on the blue (red) line construct the infinite (semi-infinite) CRW, which is called CRW-a (-b) hereafter. And the Nth cavity of a semi-infinite CRW intersects with an infinite CRW. The CTLS is situated in the node, which is the cross point of the two CRWs.

\begin{figure}[tbp]
  \includegraphics[clip=true,height=5cm,width=8.5cm]{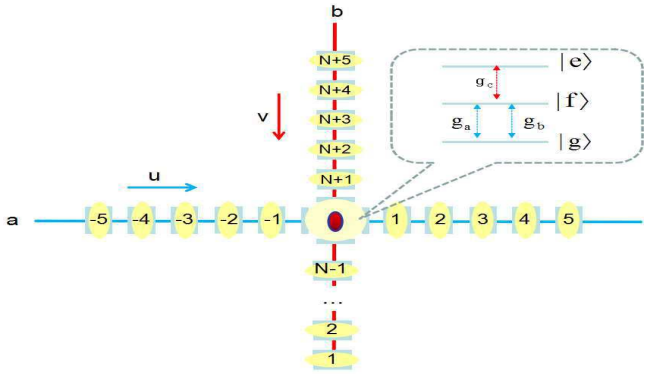}
\caption{Schematic view of quantum routing of single photons in two channels made of the Nth cavity of a semi-infinite CRW intersecting with an infinite CRW. The CTLS is characterized by $\left\vert g\right\rangle $, $\left\vert f\right\rangle $, $\left\vert e\right\rangle $, being placed at the cross point u = 0 and v = N. CRW-a (-b) couples to the CTLS through the transition $\left\vert g\right\rangle \leftrightarrow \left\vert
e\right\rangle $ with strength $g_{a}$ ($g_{b}$), transition $\left\vert e\right\rangle \leftrightarrow \left\vert
f\right\rangle $ with strength $g_{c}$. There is no interaction between the u = 0 and v = N cavities.
}
\label{fig:1}
\end{figure}

Hamilton of CTLS is below
\begin{eqnarray}
	\widehat{H_{T}}=\omega_{e}\left\vert e\right\rangle \left\langle e\right\vert + \omega _{f}\left\vert f\right\rangle \left\langle f\right\vert,
\end{eqnarray}
the CTLS's transition frequency of $\left\vert g\right\rangle\leftrightarrow \left\vert f\right\rangle$
 ($\left\vert f\right\rangle\leftrightarrow \left\vert e\right\rangle$)
 is $\omega_{f}$($\omega_{e}$).
Single photon hopes among cavities nearby, which lead to photon propagating along the single-mode cavities of the waveguides. Using tight-binding model, the two CRWs are described by the Hamiltonian
\begin{eqnarray}
	\widehat{H_{C}} = \sum^{+\infty}_{u=-\infty}[\omega_{a}a^{\dag}_{u}a_{u} - \xi_{a}(a^{\dag}_{u}a_{u+1} + a^{\dag}_{u+1}a_{u})]
+\sum^{+\infty}_{v=1}[\omega_{b}b^{\dag}_{v}b_{v} - \xi_{b}(b^{\dag}_{v}b_{v+1} + b^{\dag}_{v+1}b_{v})]+\omega_{c}c^{\dag}c.
\end{eqnarray}
Hamiltonian of interaction between atom and two CRWs is
\begin{eqnarray}
	\widehat{H_{I}}=\left\vert f\right\rangle \left\langle g\right\vert
(g_{a}a_{0}+g_{b}b_{N})+\left\vert g\right\rangle \left\langle f\right\vert (g_{a}a_{0}^{+}+g_{b}b_{N}^{+})
+\left\vert e\right\rangle \left\langle f\right\vert g_{c}c+\left\vert f\right\rangle \left\langle e\right\vert g_{c}c^{+}.
\end{eqnarray}
    The system $\widehat{H}$ consist of the two CRWs $\widehat{H_{C}}$, the CTLS $\widehat{H_{T}}$, and the interaction between atom and two CRWs $\widehat{H_{I}}$£¬
    $\widehat{H}=\widehat{H_{C}}+\widehat{H_{T}}+\widehat{H_{I}}$.
    In single excitation subspace, eigenstate of the full hamiltonian is

\begin{eqnarray}
	\left\vert E\right\rangle=(\sum^{+\infty}_{u=-\infty}U_{u}^{a}a_{u}+\sum^{+\infty}_{v=1}U_{v}^{b}b_{v})
\left\vert g,0_{a},0_{b},1_{c}\right\rangle+U_{f}\left\vert f,0_{a},0_{b},1_{c}\right\rangle+U_{e}\left\vert e,0_{a},0_{b},0_{c}\right\rangle.
\end{eqnarray}
    Fock states of CRW-a(-b) and c mode are{$0_{a}$,$1_{a}$},{$0_{b}$,$1_{b}$},{$0_{c}$,$1_{c}$}.
    The Schr\"{o}dinger Equation $\widehat{H}\left\vert E\right\rangle=E\left\vert E\right\rangle$ gives rise to a series of coupled stationary equations for all amplitudes, which is the discrete scattering equation for the single-photon propagation in the T-shaped waveguide

\begin{eqnarray}
\sum^{+\infty}_{u=-\infty}(E-\omega_{a}-\omega_{c})U_{u}^{a}=\sum^{+\infty}
_{u=-\infty}[-\xi_{a}(U_{u-1}^{a}+U_{u+1}^{a})+\delta_{u,0}(V_{a}U_0^{a}+GU
_{N}^{b})],
\end{eqnarray}
\begin{equation} \sum^{+\infty}_{v=1}(E-\omega_{b}-\omega_{c})U_{v}^{b}=\sum^{+\infty}_{v=1}
[-\xi_{b}(U_{v-1}^{b}+U_{v+1}^{b})+\delta_{v,N}(V_{b}U_{N}^{b}+GU_0^{a})],
\end{equation}
where $G\equiv g_{a}g_{b}\delta_{e}/[\delta_{e}(E-\omega_{f}-\omega_{c})-g_{c}^{2}],V_{j=a,b}\equiv g_{j}^{2}\delta_{e}/[\delta_{e}(E-\omega_{f}-\omega_{c})-g_{c}^{2}]$,the detuning between the incident photon frequency and atom transition frequency being $\delta_{e}\equiv E-\omega_{e}$.

\subsection{Single photons incident from the infinite CRW-a}
\label{sec:3}

    When the single photons incident from the CRW-a, then CRW-a contains incident light, reflected light and transmitted light, and maybe existing transfer light in CRW-b both upwords and downwords. The CTLS absorbs the photon with wavenumber k incident along the u axis onto the T-bulge shaped waveguide, transitting the CTLS from its ground state to its excited state. Since the excited state is coupled to the continua of states, the excited CTLS will emit a photon spontaneously into the propagating state of either CRW-a or CRW-b. For symmetry breaking, the wave functions between the first site and the junction of the two CRWs have standing wave form depending on N.
   \begin{equation}
U_{u}^{a}=\left\{
  \begin{array}{ll}
    e^{ik_{a}u}+re^{-ik_{a}u},u < 0; \\
    te^{ik_{a}u},u $>$ 0.
  \end{array}
\right.
\end{equation}
\begin{equation}
U_{v}^{b}=\left\{
  \begin{array}{ll}
    t^{b}e^{ik_{b}v}, & \hbox{v > N ;} \\
    A\sin \left( k_{b}N\right), & \hbox{v=1,2,3..N .}
  \end{array}
\right.
\end{equation}

The dispersion relations in the two CRWs are $E=\omega_{c}+\omega_{d}-2\xi_{d}\cos k_{d}$,{$d=a,b$}.Using continue condition, the transmission, reflection and transfer amplitude can be obtained as below, with group velocity $v_{d}=2\xi_{d}\sin k_{d}$,{$d=a,b$}.
    Nondimensionalize the the propagating amplitude(in case of $g_{a}$,$g_{b}$,$v_{a}$,$\delta_{e}$¡Ù0). In this form, the effect of N is clearer.
\begin{equation}
t^{b} =\frac{2i\sin k_{b}N}{\frac{g_{a}v_{b}}{g_{b}v_{a}}-2\frac{g_{b}}{%
g_{a}}e^{ik_{b}N}i\sin k_{b}N+i\left[ \frac{g_{c}^{2}}{\delta _{e}}-\left(
E-\omega _{s}-\omega _{c}\right) \right] \frac{v_{b}}{g_{a}g_{b}}},
\end{equation}
\begin{equation}
t =-\frac{i\left[ \frac{g_{c}^{2}}{\delta _{e}}-\left( E-\omega
_{s}-\omega _{c}\right) \right] \frac{v_{a}}{g_{a}g_{b}}-2\frac{g_{b}}{g_{a}}%
e^{ik_{b}N}i\sin k_{b}N}{\frac{g_{a}v_{b}}{g_{b}v_{a}}-2\frac{g_{b}}{g_{a}}%
e^{ik_{b}N}i\sin k_{b}N+i\left[ \frac{g_{c}^{2}}{\delta _{e}}-\left(
E-\omega _{s}-\omega _{c}\right) \right] \frac{v_{b}}{g_{a}g_{b}}},
\end{equation}
\begin{equation}
r =-\frac{\frac{g_{a}v_{b}}{g_{b}v_{a}}}{\frac{g_{a}v_{b}}{g_{b}v_{a}}-2%
\frac{g_{b}}{g_{a}}e^{ik_{b}N}i\sin k_{b}N+i\left[ \frac{g_{c}^{2}}{\delta
_{e}}-\left( E-\omega _{s}-\omega _{c}\right) \right] \frac{v_{b}}{g_{a}g_{b}%
}},
\end{equation}

   The transmission coefficients are modulus square of the transmission amplitude, $T_{b}^{a}=|t^{b}|^2, T^{a}=|t|^2, R^{a}=|r|^2$. The total possibility is conserved($T_{b}^{a}+ T^{a}+R^{a}=1$).

\begin{figure}
  \includegraphics[clip=true,height=5cm,width=8.5cm]{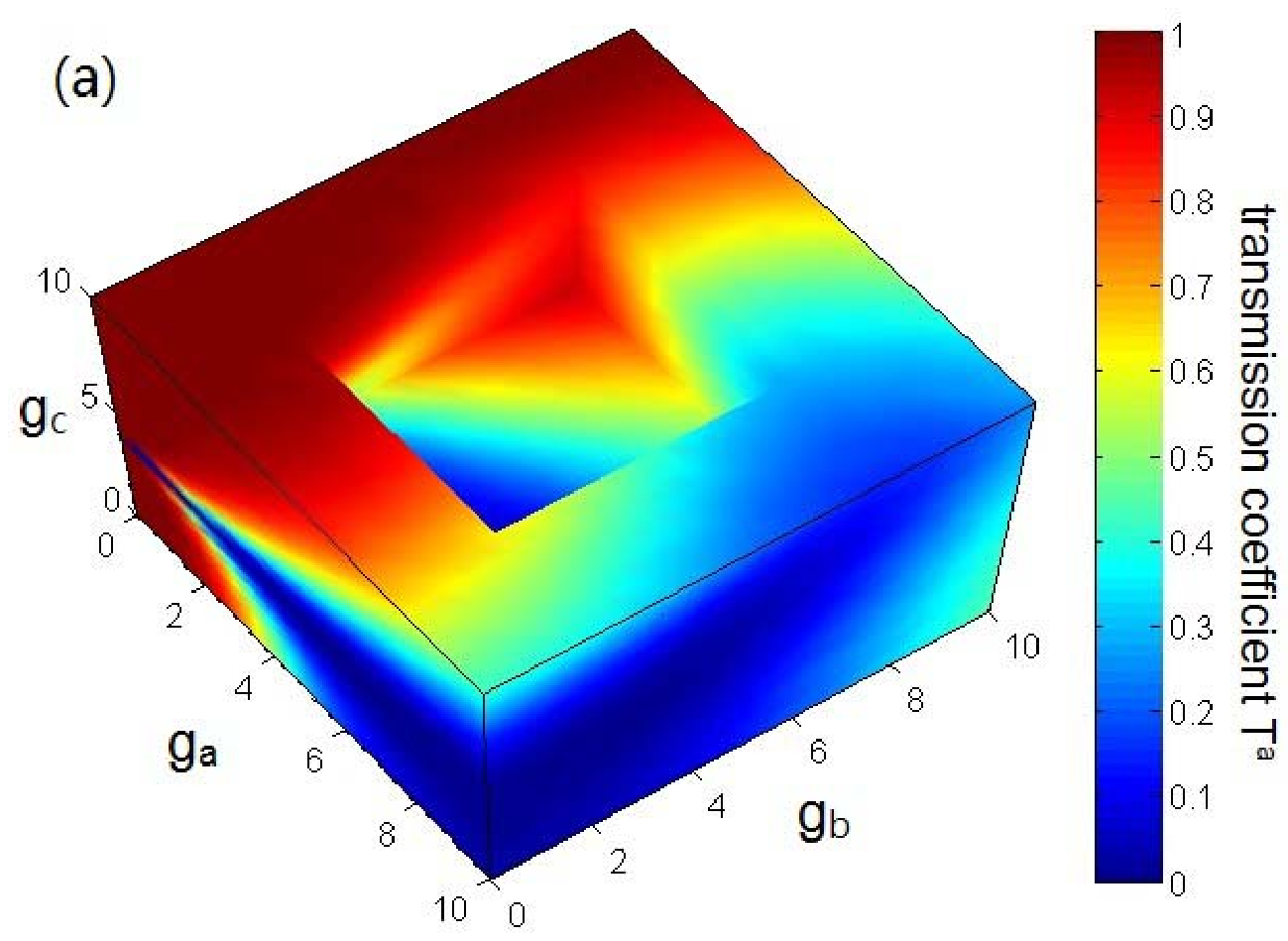}
\label{a1}       
\end{figure}

\begin{figure}
  \includegraphics[clip=true,height=5cm,width=8.5cm]{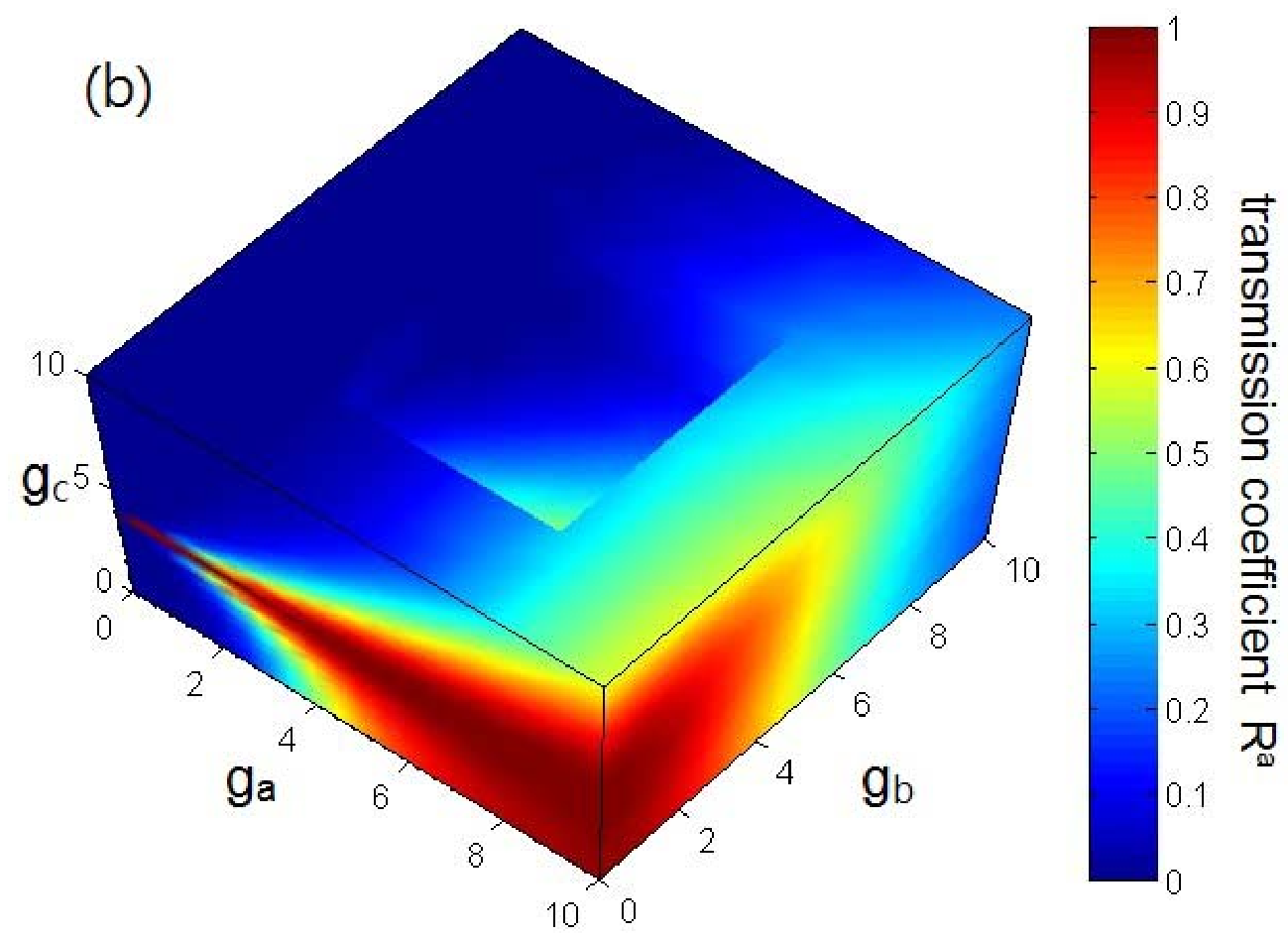}
\label{a2}       
\end{figure}

\begin{figure}
  \includegraphics[clip=true,height=5cm,width=8.5cm]{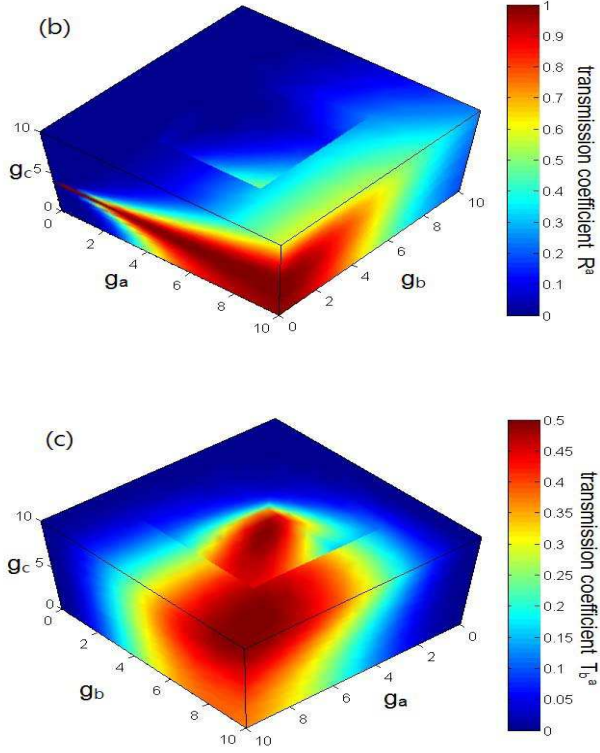}
\caption{In (a) and (b), x axis being $g_{a}$, y axis being $g_{b}$, but in (c) x axis being $g_{b}$, y axis being $g_{a}$, z axis being $g_{c}$ in all the three figures, the color shows the value gradual change from zero to one.
(a),(b) and (c) shows $T^{a}$,$R^{a}$,$T_{b}^{a}$, respectively.
Assume $\omega_{a}=\omega_{b}=\sqrt{2};\omega_{c}=\omega_{f}=3\sqrt{2};\omega_{e}=4\sqrt{2};\xi_{a}=\xi_{b}=2\xi_{0};k_{a}=k_{b}=\pi/4;N=3$. All the parameters are in units of $\xi_{0}$.
}
\label{a3}       
\end{figure}

Fig2 shows the propagation coefficients change with couple strength $g_{a}$,$g_{b}$,$g_{c}$. The centre of the 4D picture has a cuboid margin so that it's convenient to see the inner side. When $g_{a}$ is small, $T^{a}$ is tend to be one with less effect of $g_{b}$ and $g_{c}$, while $T_{b}^{a}$ and $R^{a}$ is tending to be zero. When $g_{b}$ is small, $g_{a}$ is between $\xi_{0}$ and $6\xi_{0}$, $T^{a}$ first decreases and then increases when $g_{c}$ increases, while $R^{a}$ is vice versa.
\subsection{Single photons incident from the semi-infinite CRW-b}
\label{sec:4}

    The single photons is launched into the semi-infinite CRW-b. When the traveling photon arrives at the node of the T bulge node, it is either absorbed by the CTLS or stored in the N-1 FP cavities (reflected by the boundary when $N=1$). The probability amplitudes in the CRW-a and CRW-b are given by

\begin{equation}
U_{u}^{a}=\left\{
  \begin{array}{ll}
    t_{l}^{a}e^{-ik_{a}u}, & \hbox{u < 0;} \\
    t_{r}^{a}e^{ik_{a}u}, & \hbox{u > 0.}
  \end{array}
\right.
\end{equation}
\begin{equation}
U_{v}^{b}=\left\{
  \begin{array}{ll}
    e^{-ikv}+r^{b}e^{ik_{b}v}, & \hbox{v > N ;} \\
    A\sin \left( k_{b}N\right), & \hbox{v=1,2,3..N .}
  \end{array}
\right.
\end{equation}

Using continue condition, obtaining $t_{l}^{a}=t_{l}^{a}\equiv t^{a}$, the reflection and transfer amplitude can be obtained as below
\begin{equation}
r^{b} =-\frac{\frac{g_{a}v_{b}}{g_{b}v_{a}}-2i\frac{g_{b}}{g_{a}}%
e^{-ik_{b}N}\sin k_{b}N+i\left[ \frac{g_{c}^{2}}{\delta _{e}}-\left(
E-\omega _{s}-\omega _{c}\right) \right] \frac{v_{b}}{g_{a}g_{b}}}{\frac{%
g_{a}v_{b}}{g_{b}v_{a}}-2i\frac{g_{b}}{g_{a}}e^{ik_{b}N}\sin k_{b}N+i\left[
\frac{g_{c}^{2}}{\delta _{e}}-\left( E-\omega _{s}-\omega _{c}\right) \right]
\frac{v_{b}}{g_{a}g_{b}}},
\end{equation}
\begin{equation}
t^{a} =\frac{2i\frac{v_{b}}{v_{a}}\sin k_{b}N}{\frac{g_{a}v_{b}}{g_{b}v_{a}%
}-2i\frac{g_{b}}{g_{a}}e^{ik_{b}N}\sin k_{b}N+i\left[ \frac{g_{c}^{2}}{%
\delta _{e}}-\left( E-\omega _{s}-\omega _{c}\right) \right] \frac{v_{b}}{%
g_{a}g_{b}}}
\end{equation}

    The photons incidenting from CRW-b go left and right into CRW-a with same possibilities. It's can be verified that possibility is conserved $ R^{b}+T_{a}^{b}=1,R^{b}=|r^{b}|^{2},T_{a}^{b}=2|t^{a}|^{2}$.

\begin{figure}
  \includegraphics[clip=true,height=5cm,width=8.5cm]{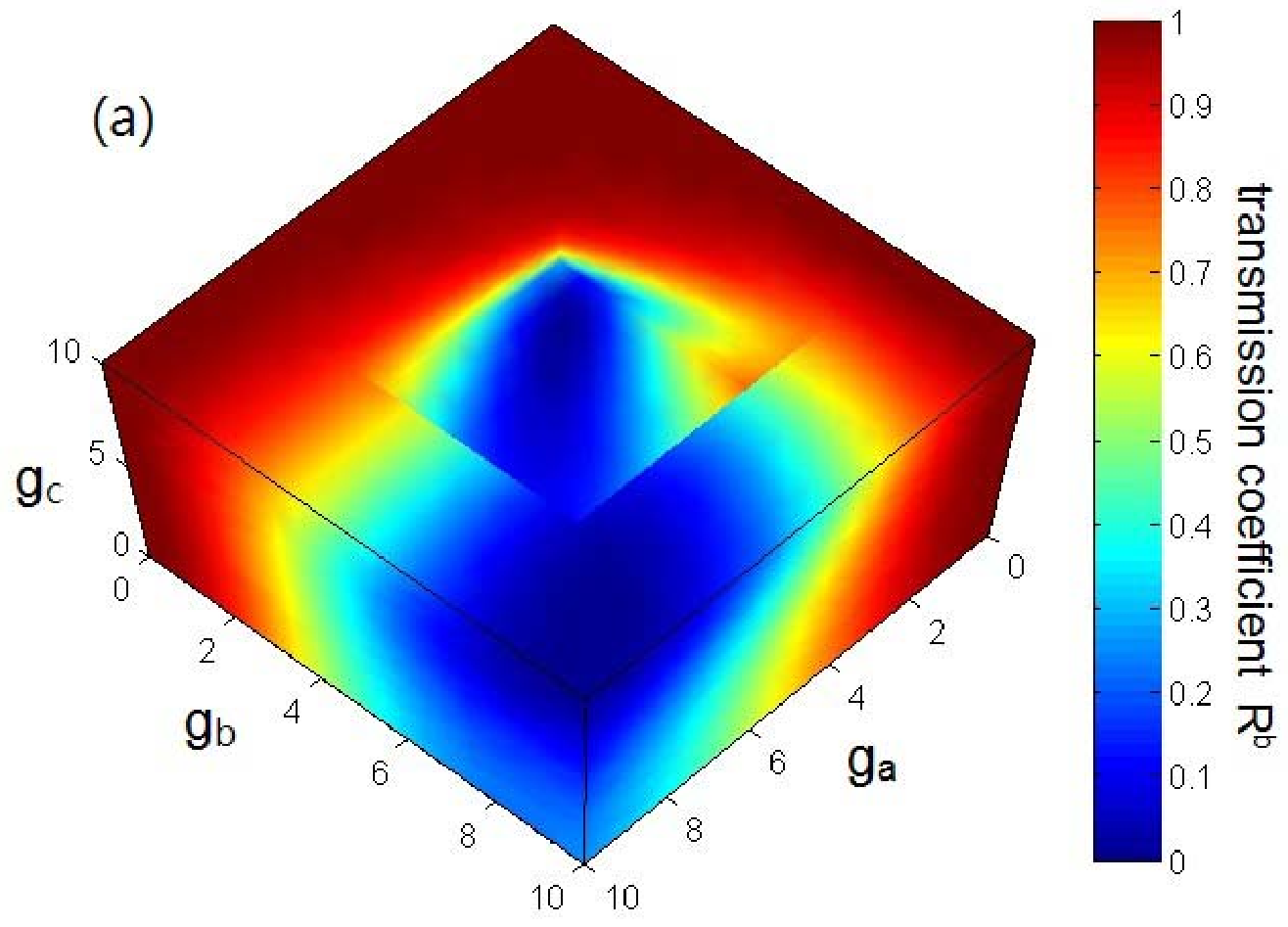}
\label{b1}       
\end{figure}

\begin{figure}
  \includegraphics[clip=true,height=5cm,width=8.5cm]{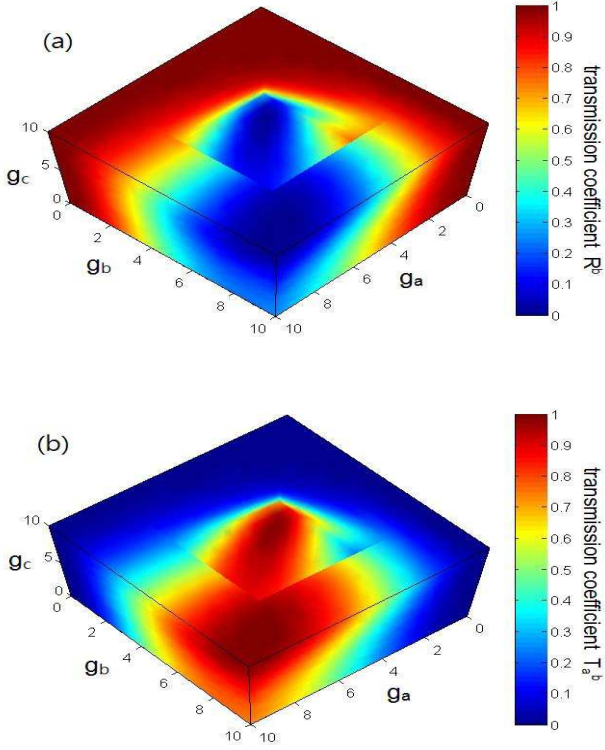}
\caption{b1(b2),x axis being $g_{b}$, y axis being $g_{a}$, z axis being $g_{c}$, the two figures shows $R^{b}$ and $T_{a}^{b}$ change with $g_{a}$ $g_{b}$ $g_{c}$.The parameters are the same as Fig 2.
}
\label{b2}       
\end{figure}

    From Fig.3, $R^{b}$ tend to be small when $g_{a}$ $g_{b}$ and $g_{c}$ to be large. But when $g_{a}$ and $g_{b}$ are small, $R^{b}$ is large no matter $g_{c}$ is large or small. $T_{a}^{b}$ is vice versa for the reason that $T_{a}^{b}=1-R^{b}$.
\section{Discussion and conclusion}
\label{Sec:5}

The effect of parameter $g_{a}$ $g_{b}$ $g_{c}$ N on the transmission coefficients of incidenting from infinte CRW-a as well as semi-infinite CRW-b are considered in this paper. When light incident from CRW-a£¨-b£©, the maximum of $T_{b}^{a}$ ($T_{a}^{b}$) would be 0.5 (1). This can be predicted for the symmetry breaking boundary. The cavity number N in the junction of the two CRWs have effect on the location of the extreme point of the curve between the transmission coefficients and couple strength.
\begin{acknowledgements}
This work is supported by the National Fundamental Research Program
of China (the 973 Program) under Grant No. 2012CB922103 and the
National Natural Science Foundation of China under Grants No.
11374095, No. 11422540, No. 11434011, No. 11575058.
\end{acknowledgements}

\end{document}